\documentclass[twocolumn]{autart}   
\usepackage{graphicx} 
\usepackage{epstopdf}
\usepackage{xcolor}
\usepackage{amsmath}
\usepackage{amssymb}
\usepackage{pgfplots}
\usepackage[list = true, hypcap = true]{subcaption}

\newtheorem{ass}[thm]{Assumption}
\newtheorem{dfn}[thm]{Definition}
\newtheorem{ntn}[thm]{Notation}

\newcommand{\IR}{\mathbb R}
\newcommand{\bmat}[1]{\begin{bmatrix} #1 \end{bmatrix}}

\begin{document}

\begin{frontmatter}

\title{Symplectic discrete-time energy-based control\\ for nonlinear mechanical systems\thanksref{footnoteinfo}} 

\thanks[footnoteinfo]{This paper was not presented at any IFAC 
meeting. Corresponding author P.~Kotyczka. Tel. +49 89 289 15663. Fax +49 89 289 15653.}

\author[first]{Paul Kotyczka}\ead{kotyczka@tum.de},
\author[first]{Tobias Thoma}\ead{tobias.thoma@tum.de}

\address[first]{Technical University of Munich, Department of Mechanical Engineering, Chair of Automatic Control, Garching, Germany} 
          
\begin{keyword}
	Sampled-data systems, discrete-time control, structure-preserving methods, symplectic integration, St{\"o}rmer-Verlet, nonlinear mechanical systems, passivity-based control, energy shaping, computed torque.
\end{keyword}

\begin{abstract}
	In this article we present a novel discrete-time design approach which reduces the deteriorating effects of sampling on stability and performance in digitally controlled nonlinear mechanical systems. The method is motivated by recent results for linear systems, where feedback imposes closed-loop behavior that exactly represents the symplectic discretization of a desired target system. In the nonlinear case, both the second order accurate representation of the sampling process and the definition of the target dynamics stem from the application of the implicit midpoint rule. The implicit nature of the resulting state feedback requires the numerical solution of an in general nonlinear system of algebraic equations in every sampling interval. For an implementation with pure position feedback, the velocities/momenta have to be approximated in the sampling instants, which gives a clear interpretation of our approach in terms of the St{\"o}rmer-Verlet integration scheme on a staggered grid. We present discrete-time versions of impedance or energy shaping plus damping injection control as well as computed torque tracking control. Both the Hamiltonian and the Lagrangian perspective are adopted. Besides a linear example to introduce the concept, the simulations with a planar two link robot model illustrate the performance and stability gain compared to the discrete implementations of continuous-time control laws. A short analysis of computation times shows the real-time capability of our method.
\end{abstract}

\end{frontmatter}

\date{9 June 2020}

\section{Introduction}

Passivity-based methods are an important framework for the feedback control of mechanical systems, in particular in robotics: From potential-based PD control \cite{takegaki1981new}, over a generalized Euler-Lagrangian approach \cite{ortega1998passivity}, designs for underactuated systems from both the Lagrangian \cite{woolsey2004controlled} and the Hamiltonian \cite{ortega2002stabilization} perspective, up to a unified approach for different control tasks in flexible joint robot control \cite{albuschaeffer2007unified}, \emph{energy shaping} is a well-established control design argument. Also \emph{computed torque}, i.e., feedback linearization (for this and other familiar motion control designs in robotics, see e.g., \cite{canudas1996theory,siciliano2016handbook,spong2008robot}) can be parametrized such that the tracking error follows a second order ODE with shaped mechanical structure.

A direct discrete implementation of the continuously derived control laws in digital, sampled (robot) control systems leads to performance degradation and instability, see e.g. \cite{chen1987frequency}, and the design of discrete-time controllers based on discrete models is the appropriate answer. Early works are \cite{neuman1985discrete}, where integrals of motion are conserved and dynamics is approximated with the trapezoidal rule, or \cite{nicosia1989discrete} with the discretization of the action integral using Euler approximations of the velocities. The above mentioned trapezoidal rule -- which is conjugate to the symplectic implicit midpoint rule, see \cite{hairer2006geometric}, Section VI.8 -- is also popular for the high accuracy discrete implementation of continuously computed robot controllers, see \cite{siciliano2016handbook}, Section 6.9.2.

Symplectic integration schemes, see e.g. \cite{hairer2006geometric}, are derived with the goal of preserving the symplecticity (area/volume preservation in phase space) of the flow of a Hamiltonian system in discrete time. This property implies preservation of invariants of motion, and an excellent long time behavior in the simulation of conservative systems due to the conservation of a modified Hamiltonian. They also go along with a meaningful approximation of transferred power, which is instrumental in defining discrete-time port-Hamiltonian systems \cite{kotyczka2019discreteSCL}. Symplectic integration is also successfully used in the solution of optimal control problems, see e.g., \cite{peng2012symplectic}. Symplectic schemes are closely related to variational integrators, which are a dual approach for the structure-preserving integration of mechanical systems from the Lagrangian perspective. In particular, symplecticity of the numerical scheme can be enforced by the construction of the variational integrator \cite{lew2004variational}. Variational integrators are as well widely used for the solution of optimal control problems \cite{oberbloebaum2010discrete}.

Recently, discrete-time eigenvalue assignment based on the definition of a target system using the -- symplectic -- implicit midpoint rule was proposed in \cite{kotyczka2020discrete}. Advantages compared to the implementation of continuous controllers are unconditional stability w.r.t. the sampling time (within the limits of the sampling theorem) and the possibility to impose conservative discrete-time behavior, which corresponds to energy shaping. Conservative (target) dynamics and long time frames are conditions, where symplectic integration schemes for control show particular advantages compared to other methods \cite{chyba2009role}.

The main interest of the presented research is the systematic formulation of discrete-time passivity-based control techniques for mechanical, in particular robotic, systems based on symplectic integration.  Energy shaping is the central argument of passivity-based controls, and symplectic numerical schemes guarantee the conservation of a modified Hamiltonian. They are therefore appropriate to endow the closed-loop system with the desired conservativeness in the energy shaping step. Our approach is based on a second order accurate description of the sampled nonlinear mechanical control system using the implicit midpoint rule. Accordingly, the target dynamics for control (in the first step energy shaping) is defined based on this integration scheme. Comparison of open- and closed-loop system yields the control law in implicit form. By the solution of a system of in general nonlinear algebraic equations, the control input, which then only depends on coordinates and velocities (momenta) at the given sampling instant, can be computed. In general, only position measurements are available for feedback, and velocities (momenta) must be numerically computed. Reconstructing them using the trapezoidal rule, and therefore enabling pure position feedback, leads to a beautiful interpretation of the resulting discrete-time system: It can be considered the discretization of the underlying continuous-time target systems with the -- symplectic -- St{\"o}rmer-Verlet scheme on a staggered grid. 
The application of the approach to modified target systems -- with shaped kinetic energy, gyroscopic and dissipative forces -- is a straightforward extension.
In order to concentrate on a clear introduction to this new approach, we restrict ourselves to the fully actuated case, i.e., all mechanical degrees of freedom are assumed controlled. 

The article is organized as follows. In Section \ref{sec:2} we give background information on the Hamiltonian and Lagrangian representation of mechanical systems, the considered (not only) passivity-based control techniques and on symplectic integration. Section \ref{sec:3} contains the main results, which are based on the implicit midpoint rule (or one-stage Gauss collocation method). The representation of the sampling process is followed by the introduction of discrete-time energy shaping control from the Hamiltonian point of view and the inclusion of dissipation an/or gyroscopic forces in the same frame. Our approach is then formulated as pure position feedback. Finally, the dual Lagrangian perspective is considered, which is more usual when computed torque -- which also can be interpreted in terms of total energy shaping for the error system -- is applied. The interpretation in terms of the St{\"o}rmer-Verlet scheme brings us back to symplectic integration. The three simulation examples -- for the control tasks impedance, set point and trajectory tracking control -- in Section \ref{sec:4} illustrate the quality of our approach and the performance gain compared to quasi-continuous control. The analysis of the computation times for the implicitly given control laws confirms the applicability of our approach on industrial hardware. We close the article with remarks and an outlook to further works in Section \ref{sec:5}.

\section{Preliminaries}
\label{sec:2}
We recall the Hamiltonian and Lagrangian representations of mechanical systems, some popular control techniques from robotics, and we give a brief introduction to symplectic integration and structure preservation.

\subsection{Models of mechanical systems}
The dynamics of a simple\footnote{The energy is split in potential and kinetic energy.}, lossless mechanical system can be written in canonical \emph{Hamiltonian} form 
\begin{equation}
	\label{eq:20.30.10}
	\bmat{\dot q(t)\\ \dot p(t)}
	= 
	\bmat{0 & I\\ - I & 0}
	\bmat{\nabla_q H(q(t),p(t))\\ \nabla_p H(q(t),p(t))}
	+
	\bmat{0 \\ I}
	u(t)
	\end{equation}
with 
\begin{equation}
	\label{eq:20.30.20}
	H(q(t),p(t)) = \frac{1}{2} p^T(t) M^{-1}(q(t)) p(t) + V(q(t))
\end{equation}
the Hamiltonian. $V: \IR^n \rightarrow \IR$ denotes the potential energy and the first term represents the kinetic energy $T: \IR^n \times \IR^n \rightarrow \IR$ with  $M: \IR^n \rightarrow \IR^{n\times n}$ the positive definite, symmetric mass or inertia matrix. $q,p \in \IR^n$ are the generalized coordinates and momenta, respectively\footnote{In this article, we do not assume a differential-geometric perspective, therefore we identify the spaces of coordinates, velocities and momenta with $\IR^n$, instead of stressing their nature on the configuration manifold $Q$, the tangent and the co-tangent bundle $TQ$ and $T^*Q$, respectively.}. $u(t) \in \IR ^n$ denotes the vector of input forces/torques for the \emph{fully actuated mechanical system}. Throughout the paper we will use the representation
\begin{subequations}
\label{eq:3.010}
	\begin{align}
	\label{eq:3.010a}
	\dot q(t) &= M^{-1}(q(t)) p(t)\\
	\label{eq:3.010b}
	\dot p(t) &= - \nabla_q H (q(t),p(t)) + u(t).
	\end{align}
\end{subequations}
The equivalent \emph{second order} representation is based on the Lagrangian
\begin{equation}
	\label{eq:20.30.30}
	L(q(t),\dot q(t)) = \frac{1}{2} \dot q(t)^T M(q(t)) \dot q (t) - V(q(t)).
\end{equation}
The first term is now the mechanical (co-)energy, expressed in the generalized velocities $\dot q \in \IR^n$. Hamiltonian and Lagrangian representation are coupled via the \emph{Legendre transform} with $H(q,p) + L(q,\dot q) = p^T \dot q$. Evaluation of the Euler-Lagrange equations yields then
\begin{equation}
	\label{eq:20.30.40}
	M(q(t))\ddot{q}(t) + C(q(t),\dot{q}(t)) \dot{q}(t) + \nabla V(q(t)) = u(t),
\end{equation}
where the Coriolis matrix $C: \IR^n\times \IR^n \rightarrow \IR^{n \times n}$ and the mass matrix are related via the property that $\dot M - 2 C$ is skew-symmetric. With the vector of velocities $v=\dot q$, the first order formulation of \eqref{eq:20.30.40} is 
\begin{subequations}
\label{eq:20.30.41}
\begin{align}
	\dot q(t) &= v(t)\\
	\dot v(t) &= f(q(t), v(t)) + M^{-1}(q(t)) u(t),
\end{align}
\end{subequations}
where $f = - M^{-1} C - M^{-1} \nabla V$.

\begin{rem}
	Neglecting dissipation in the open-loop model is not a severe restriction, as friction in a modern motion control structure is typically counteracted via additional actions, see e.g. \cite{bona2005friction} for an overview. Moreover, the results in the following section can be extended to the dissipative case in a straightforward manner. Their representation for the conservative case is, however, more compact.
\end{rem}

\subsection{Control of mechanical systems}
\label{subsec:control-mech-sys}
We summarize some popular methods for the control of fully actuated mechanical systems in continuous time, first from the Hamiltonian, then from the Lagrangian perspective. Discrete-time versions will be presented in the following section.

\subsubsection{Passivity-based control, energy shaping}
The idea of \emph{Interconnection and Damping Assignment Passivity-Based Control} (IDA-PBC) is straightforward, see e.g. \cite{ortega2002stabilization}. By state feedback, the mechanical control system \eqref{eq:20.30.10} is endowed with a new (artificial) Hamiltonian structure
\begin{equation}
	\label{eq:20.30.50}
	\bmat{\dot q(t)\\ \dot p(t)}
	= 
	\bmat{0 & J_1(q)\\ - J_1^T(q) & F_2(q(t), p(t))}
	\bmat{\nabla_q H_d(q(t),p(t))\\ \nabla_p H_d(q(t),p(t))}
\end{equation}
where 
\begin{equation}
	\label{eq:20.30.60}
	H_d(q(t),p(t)) = \frac{1}{2} p(t)^T M_d^{-1}(q(t)) p(t) + V_d(q(t))
\end{equation}
is the \emph{shaped Hamiltonian} with $M_d(q) = M_d^T(q) > 0$ the artificial inertia matrix and $V_d(q)$ the shaped potential energy with a strict minimum in the desired equilibrium configuration $q^*$. To guarantee that the first equations of \eqref{eq:20.30.10} and \eqref{eq:20.30.50} match, $J_1(q) = M^{-1}(q) M_d(q)$ must be chosen. The matrix 
\begin{equation}
	F_2(q(t), p(t)) = J_2(q(t), p(t)) - R_2(q(t), p(t))
\end{equation}
with $J_2(\cdot, \cdot) = - J_2^T(\cdot, \cdot)$ and $R_2(\cdot, \cdot) = R_2^T(\cdot, \cdot) \geq 0$ represents additional gyroscopic terms and dissipation, which enforces asymptotic stability of the desired equilibrium\footnote{If the damping is \emph{pervasive}, for which a controllability type criterion can be formulated \cite{mueller1972stability}, asymptotic stability of (linear) mechanical systems can be easily proven also in the case of a semi-definite damping matrix.}. The continuous-time control law is simply obtained by comparison of the second lines of \eqref{eq:20.30.10} and \eqref{eq:20.30.50} (for brevity, we omit all arguments):
\begin{equation}
	\label{eq:20.30.51}
	u = \nabla_q H - M_d M^{-1} \nabla_q H_d + F_2 M_d^{-1} p.
\end{equation}
Impedance control, where the ``controller attempts to implement a dynamic relation between manipulator variables such as end-point position and force'' \cite{hogan1984impedance}, can, for example, be expressed as an energy shaping problem. Potential energy and dissipation structure are assigned according to the desired characteristics of the virtual spring-damper system. The unifying perspective of passivity-based control is extensively presented in \cite{albuschaeffer2007unified} for the (underactuated) case of series elastic manipulators with torque feedback.

\begin{rem}
	Obviously, for fully actuated mechanical systems, the solution of the energy shaping problem is trivial. It is more difficult in the underactuated case \cite{viola2007total}, where the design parameters must satisfy matching equations for potential and kinetic energy and dissipation.
\end{rem} 

In this paper, we will first deal with the special case of only potential energy shaping, i.e., $M_d(q) = M(q)$ and $J_1(q) = I$. The target system can then be written
\begin{subequations}
	\label{eq:3.070}
	\begin{align}
	\label{eq:3.070a}		
	\dot q(t) &= M^{-1}(q(t)) p(t)\\
	\label{eq:3.070b}
	\dot p(t) &= -\nabla_q H_d(q(t),p(t)),
	\end{align}
\end{subequations}
and the control law boils down to
\begin{equation}
	\label{eq:20.30.52}
	u(t) = \nabla V(q(t)) - \nabla V_d(q(t)).
\end{equation}
The results will then be extended to target systems of the general form
\begin{subequations}
	\label{eq:3.072}
	\begin{align}
	\label{eq:3.072a}		
	\dot q(t) &= M^{-1}(q(t)) p(t)\\
	\label{eq:3.072b}
	\dot p(t) &= b_d(q(t),p(t)),
	\end{align}
\end{subequations}
and the control law
\begin{equation}
	\label{eq:20.30.54}
	u(t) = \nabla_q H(q(t), p(t)) + b_d(q(t), p(t)),
\end{equation}
which includes \eqref{eq:20.30.50} and \eqref{eq:20.30.51} as a particular case.

\begin{rem}
	For \emph{passivity-based control}, without the port-Hamiltonian perspective, but with a focus on trajectory tracking, we refer for example to \cite{canudas1996theory}, Section 2.3.3, or \cite{spong2008robot}, Section 11.4.2.
\end{rem}

\subsubsection{PD control with gravity compensation}
\label{subsec:PD}
A standard robot controller for set point stabilization in joint space, see e.g. \cite{canudas1996theory}, Section 2.2.3, or \cite{spong2008robot}, Section 11.2, is
	\begin{equation}
		\label{eq:4.210}
		u(t) = \nabla V(q(t)) - D\dot{q}(t) - K(q(t) - q_d).
	\end{equation}
$q_d \in \IR^n$ is the desired configuration/position vector and $K$, $D$ are constant, symmetric positive definite matrices. The control law \eqref{eq:4.210}, applied to \eqref{eq:20.30.40}, generates the target system
	\begin{equation}
		\label{eq:4.220}
		M(q(t))\ddot{q}(t) + (C(q(t),\dot{q}(t))+D)\dot{q}(t) + K(q(t)-q_d) = 0.
	\end{equation}
It represents a mechanical system with the quadratic potential energy $\frac{1}{2} (q - q_d)^T K (q- q_d)$ of a virtual spring and additional velocity-dependent damping forces.

\subsubsection{Computed torque}
\label{subsec:computed-torque}
In extension to \eqref{eq:4.210}, the \emph{computed torque} (feedback linearization) control law, see e.g. \cite{canudas1996theory}, Section 2.3.1, or \cite{spong2008robot}, Section 11.3,
	\begin{equation}
		\label{eq:4.230}
		u = C\dot{q} + \nabla V + M\ddot{q}_d + MM_d^{-1}(-Ke-D\dot{e})
	\end{equation}
compensates both potential and Coriolis forces in \eqref{eq:20.30.40}. With $e(t) = q(t) - q_d(t)$ the tracking error for a given trajectory $q_d: [0,\infty) \rightarrow \IR^n$ and constant, symmetric positive definite matrices $M_d$, $K$ and $D$, the resulting dynamical system for the tracking error
	\begin{equation}
		\label{eq:4.240}
		M_d\ddot{e}(t) + D\dot{e}(t) + Ke(t) = 0
	\end{equation}
has a linear mechanical structure, which implies asymptotic trajectory tracking, $\lim_{t\rightarrow \infty} e(t) = 0$.

\begin{rem}
	In the context of computed torque methods, it is interesting to note that the trajectory tracking problem for rigid and flexible joint robots in continuous time without velocity measurement has been solved in \cite{loria1995tracking} using a filtered position feedback.
\end{rem}

In the following section we will also use the first order representation of the target systems \eqref{eq:4.220} and \eqref{eq:4.240},
\begin{subequations}
\label{eq:20.30.43}
\begin{align}
	\dot q(t) &= v(t)\\
	\dot v(t) &= f_d(q(t), v(t), t),
\end{align}
\end{subequations}
where $f_d$ is time-varying in the case of trajectory tracking control.

\subsection{Symplectic integration}
\label{subsec:2.1}
A particular interest in the numerical integration of Hamiltonian systems is the preservation of structural properties. \emph{Symplecticity} refers to area/volume preservation in phase space and implies the \emph{conservation} of a modified Hamiltonian in the undamped case, which can be verified by \emph{backward error analysis} \cite{hairer2006geometric}. From the Lagrangian point of view, the appropriate discretization of the Lagrangian and the action functional leads to discrete variational principles and \emph{variational integrators}, which also represent symplectic schemes \cite{lew2004variational,oberbloebaum2010discrete}. Two classical symplectic methods play a key role in this article.

\subsubsection{Implicit midpoint rule}
The numerical solution of a dynamical system $\dot x(t) = f(x(t), t)$ with the \emph{implicit midpoint rule} or \emph{one-stage Gauss collocation scheme} is computed by
\begin{equation}
	\label{eq:20.30.90}
	x_{k+1} = x_k + h f(x_{k+\frac{1}{2}}, t_{k+\frac{1}{2}}),
\end{equation}
where $t_{k+\frac{1}{2}} = t_k + \frac{h}{2}$, and the stage value $x_{k+\frac{1}{2}}$ is determined from the half implicit Euler step
\begin{equation}
	\label{eq:20.30.100}
	x_{k+\frac{1}{2}} = x_k + \frac{h}{2} f(x_{k+\frac{1}{2}}, t_{k+\frac{1}{2}}).
\end{equation}
Combining both equations, we obtain the simple formula
\begin{equation}
	\label{eq:20.30.110}
	x_{k+\frac{1}{2}} = \frac{x_k + x_{k+1}}{2},
\end{equation}
which gives the name to the method. The implicit midpoint rule is one of the simplest symplectic schemes and has an approximation order of 2.

\subsubsection{St{\"o}rmer-Verlet scheme}
This second order integration scheme goes back to \cite{stormer1907trajectoires} and \cite{verlet1967computer}, where it was first used for numerical integration of models in celestial mechanics and molecule dynamics. It applies to partitioned systems of the form $\dot q(t) = v(t)$, $\dot v(t) = f(q(t))$. It can be written (among other representations) as a two-step scheme
\begin{equation}
	\label{eq:20.30.70}
	q_{k+1} - 2 q_k + q_{k-1} = h^2 f(q_k)
\end{equation}
or in a one-step formulation on \emph{staggered grids} for position and velocity,
\begin{equation}
	\label{eq:20.30.80}
	\begin{split}
		v_{k+\frac{1}{2}} &= v_{k-\frac{1}{2}} + h f(q_k)\\
		q_{k+\frac{1}{2}} &= q_k + h v_{k+\frac{1}{2}},
	\end{split}
\end{equation}
which is the ``computationally most economic implementation'' of the St{\"o}rmer-Verlet scheme \cite{hairer2003geometric}.

\section{Main results}
\label{sec:3}
\begin{figure}
\begin{center}
	\includegraphics[width=8.4cm]{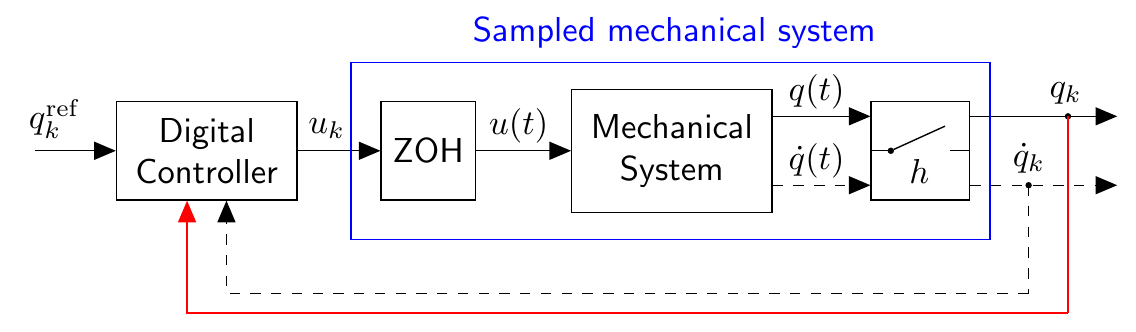}    
	\caption{Sampled mechanical control system. The blue box represents the discrete-time mechanical system, for which we will give a second order approximation. The dashed feedback indicates that velocities are typically not directly measured. The main result of this paper is symplectic feedback control design based on exclusively the (red) position feedback.}  
	\label{fig:sampled-control-loop}                                 
\end{center}  
\end{figure}

To introduce our control method, we start with the Hamiltonian representation, as it immediately visualizes the direct action of the inputs on the momenta, and the relation between velocities and momenta, which must be preserved in closed loop. The system is considered as part of a digital control loop as sketched in Fig. \ref{fig:sampled-control-loop}.

\subsection{Representation of the sampling process}
We propose a discrete-time control model, which represents the sampling process according to Fig. \ref{fig:sampled-control-loop} with second order accuracy in the time step $h$. The model has a particularly simple structure in the update equation for the coordinates $q$, which makes it well suited for the proposed symplectic discrete-time control design.

\begin{ass}
	\label{ass:01}
	We consider piecewise constant inputs $u(t) = u_k$, $t_k \leq t < t_{k+1}$, and a constant sampling time $h$, i.e., $t_{k} = k h$, $k \in \mathbb N_0$. The position vector $q_k$ is assumed measurable in every sampling instant. The initial assumption that $\dot q_k$ and therefore $p_k$ is available, will be removed for the main result of the paper.
\end{ass}

\begin{ntn}
	The inverse mass matrices are denoted
		$\bar M(q):= M^{-1}(q)$ and
		$\bar M_d(q):= M^{-1}_d(q)$.
\end{ntn}

The discrete-time control model will be based on the implicit midpoint rule or one-stage Gauss collocation.
\begin{ntn}
	We write the stage values
	\begin{equation}
		\label{eq:3.025}
		q_{k+\frac{1}{2}}:= \frac{q_k + q_{k+1}}{2} \quad \text{and} \quad  p_{k+\frac{1}{2}} := \frac{p_k + p_{k+1}}{2}.
	\end{equation}
\end{ntn}

\begin{thm}
	\label{thm:01}
	The implicit discrete-time control model
	\begin{subequations}
	\label{eq:3.030}
	\begin{align}
		\label{eq:3.030a}
		q_{k+1} &= q_k + h \bar M(q_{k+\frac{1}{2}}) p_{k+\frac{1}{2}}\\ 
		\label{eq:3.030b}
		p_{k+1} &= p_k - h \nabla_q H(q_{k+\frac{1}{2}}, p_{k+\frac{1}{2}}) + h u_k	
	\end{align}
	\end{subequations}
	is a second order approximation of the mechanical system \eqref{eq:3.010} under piecewise constant input $u(t)$ (zero order hold) and with sampling time $h$.
\end{thm}

\begin{pf}
	The exact solutions of Eqs. \eqref{eq:3.010a} and \eqref{eq:3.010b} at the end of a sampling interval $[t_k, t_{k+1}]$, under the assumption of piecewise constant inputs, are
	\begin{subequations}
		\label{eq:3.040}
		\begin{align}
			\label{eq:3.040a}
			q_{k+1} &=
			q_{k} + \int_{t_k}^{t_{k+1}} \bar M(q(\tau)) p(\tau)\, d\tau \\
			\label{eq:3.040b}
			p_{k+1} &=
			p_{k} - \int_{t_k}^{t_{k+1}} \nabla_q H(q(\tau), p(\tau))\, d\tau
			+ h u_k.
		\end{align}
	\end{subequations}
	The corresponding terms in \eqref{eq:3.030} are the numerical approximations of the integrals in \eqref{eq:3.040} with the implicit midpoint rule or one-stage Gauss quadrature. This quadrature rule is known to be exact up to polynomial degree 2, see \cite{hairer2006geometric}, Section II.1.3. Therefore, \eqref{eq:3.030} approximates \eqref{eq:3.040} with an error $o(h^2)$, i.e., of order greater than 2 in $h$.
    \hfill ${ }$ \qed
\end{pf}

In what follows, we consider \eqref{eq:3.030} as the discrete-time control model, knowing (a) that it approximates the sampling process with second order accuracy and (b) that the approximation error oscillates around zero due to the symplecticity of the integration scheme.

\subsection{Potential energy shaping}
Our control approach is first presented for conservative desired closed-loop systems with only modified \emph{potential} energy. In that scenario the symplecticity of the implicit midpoint rule is exploited to guarantee the required conservativeness of the discrete-time target system.

To achieve different control tasks like set point, impedance control or trajectory tracking, however, the inclusion of dissipation is necessary. Also additional gyroscopic forces and/or the shaping of the kinetic energy represent interesting degrees of freedom to impose desired closed-loop dynamics. Therefore, a generalized version of the \emph{implicit symplectic control design} approach is presented further below.

The discrete-time update equation for the generalized coordinates \eqref{eq:3.030a} is not affected by the input $u_k$, other than the update equation for the momenta \eqref{eq:3.030b}, which can be \emph{arbitrarily} modified by control due to the assumption of \emph{full actuation}. This fact suggests the following definition of the \emph{discrete-time target system}.

\begin{dfn}
	The target system for discrete-time potential energy shaping is given by
	\begin{subequations}
	\label{eq:3.060}
	\begin{align}
	\label{eq:3.060a}
		q_{k+1} &= q_{k} + h \bar M(q_{k+\frac{1}{2}}) p_{k+\frac{1}{2}}\\
		\label{eq:3.060b}
		p_{k+1} &= p_{k} - h \nabla_q H_d(q_{k+\frac{1}{2}}, p_{k+\frac{1}{2}}),
	\end{align}
	\end{subequations}
	with Hamiltonian $H_d(q,p) = V_d(q) + \frac{1}{2} p^T \bar M(q) p$. The shaped potential energy has a strict minimum at the desired closed-loop configuration: $q^* = \arg \min_q V_d(q)$.
\end{dfn}	

The so-defined target system is the implicit midpoint rule discretization of \eqref{eq:3.070}, and the following theorem describes the transition from \eqref{eq:3.030} to \eqref{eq:3.060}.

\begin{thm}
	\label{thm:02}
	Take $q_{k+\frac{1}{2}}$ and $p_{k+\frac{1}{2}}$ as solutions of the system of equations
	\begin{subequations}
		\label{eq:3.062}
		\begin{align}
			\label{eq:3.062a}
			M(q_{k+\frac{1}{2}}) (q_{k+\frac{1}{2}} - q_{k}) &= \frac{h}{2} p_{k+\frac{1}{2}}\\
			\label{eq:3.062b}
			p_{k+\frac{1}{2}} - p_k &= - \frac{h}{2} \nabla_q H_d(q_{k+\frac{1}{2}}, p_{k+\frac{1}{2}}).
		\end{align}
	\end{subequations}
	The (implicit) discrete-time potential energy shaping state feedback control
	\begin{equation}
		\label{eq:3.090}
		u_k =  \nabla V(q_{k+\frac{1}{2}}) - \nabla V_d(q_{k+\frac{1}{2}})
	\end{equation}
	imposes the target dynamics \eqref{eq:3.070} on the sampled version of the open-loop system \eqref{eq:20.30.10} with (a) an error of order $o(h^2)$, as opposed (b) to an error $o(h)$ under the discrete-time implementation of the continuous-time control law \eqref{eq:20.30.52}, $u_k = \nabla V(q_k) - \nabla V_d(q_k)$.
\end{thm}

\begin{pf}
	(a) With the definition of the stage values $q_{k+\frac{1}{2}}$ and $p_{k+\frac{1}{2}}$ according to \eqref{eq:3.025}, the system of equations \eqref{eq:3.062} is exactly the desired target dynamics \eqref{eq:3.060}, which is a second order approximation of \eqref{eq:3.070}. The implicit control law \eqref{eq:3.090} transforms  \eqref{eq:3.030b}, which is also a second order approximation of the continuous momentum equation \eqref{eq:3.010b}, into \eqref{eq:3.060b}. For the proof of statement (b), we obtain the order $o(h)$ of the error between the two resulting closed-loop systems by Taylor series expansion around a common expansion point. \hfill \quad \qed
\end{pf}

\begin{rem}
	It is easy to see that \eqref{eq:3.062} represents the computation of the unknown stage values with an implicit Euler step of size $\frac{h}{2}$, as introduced in \eqref{eq:20.30.100}.
\end{rem}

\subsection{Total energy shaping}
A conservative system as \eqref{eq:3.070} is not the target system for passivity-based control if the control goal is the asymptotic stabilization of a set point or the error along a desired trajectory. Also may only potential energy shaping be insufficient to modify the dynamic behavior in the desired way. Therefore, we now formulate the implicit discrete-time control design for the general target system \eqref{eq:3.072}, knowing that the control law \eqref{eq:20.30.54} does the job in continuous time. We first apply the implicit midpoint rule to \eqref{eq:3.072}.

\begin{dfn}
	\label{dfn:target-discrete-generalized}
	A generalized discrete-time target system, which includes the case \eqref{eq:20.30.50} of total energy shaping plus damping injection and gyroscopic forces, has the form
	\begin{subequations}
		\label{eq:3.110}
		\begin{align}
		\label{eq:3.110a}
			q_{k+1}  &= q_{k} + h \bar M(q_{k+\frac{1}{2}}) p_{k+\frac{1}{2}}\\
			\label{eq:3.110b}
			p_{k+1} &= p_{k}  + h b_d(q_{k+\frac{1}{2}}, p_{k+\frac{1}{2}}).
		\end{align}
		\end{subequations}	
\end{dfn}
We can determine the discrete-time control law to bring the control model \eqref{eq:3.030} into this form in complete accordance with Theorem \ref{thm:02}.
\begin{cor}
	\label{cor:01}
	The discrete-time implicit state feedback 
	\begin{multline}
		\label{eq:3.111}
		u_k = \nabla_q H(q_{k+\frac{1}{2}}, p_{k+\frac{1}{2}}) + b_d(q_{k+\frac{1}{2}}, p_{k+\frac{1}{2}})
	\end{multline}
	with $q_{k+\frac{1}{2}}$ and $p_{k+\frac{1}{2}}$ solutions of the nonlinear system of equations
	\begin{subequations}
		\label{eq:3.112}
		\begin{align}
			\label{eq:3.112a}
			M(q_{k+\frac{1}{2}}) (q_{k+\frac{1}{2}} - q_{k}) &= \frac{h}{2} p_{k+\frac{1}{2}}\\
			\label{eq:3.112b}
			p_{k+\frac{1}{2}} - p_k &= \frac{h}{2} b_d(q_{k+\frac{1}{2}}, p_{k+\frac{1}{2}})
		\end{align}
	\end{subequations}
	imposes the discrete-time version \eqref{eq:3.110} of the desired continuous-time dynamics \eqref{eq:3.072} with an error of order $o(h^2)$ on the sampled mechanical system \eqref{eq:3.010}.
\end{cor}

\begin{pf}
	The proof is parallel to the proof of Theorem \ref{thm:02}. Again, \eqref{eq:3.112} in conjunction with the definition of the stage values \eqref{eq:3.025}, represents the target system \eqref{eq:3.110}, which is generated from \eqref{eq:3.030} by the application of the control law \eqref{eq:3.111}. Both discrete-time systems are second order approximations of the sampled dynamics \eqref{eq:3.040} and the continuous-time target system \eqref{eq:3.072}.
	\hfill \quad \qed
\end{pf}

\begin{rem}
	Imposing a (strongly) damped target behavior with short-time desired transients is not the typical scenario for symplectic integration to showcase its advantages. Nevertheless, a structured control design, which separates energy shaping from the injection of damping and gyroscopic terms, definitely motivates the use of discretization schemes with conservation properties.
\end{rem}

\subsection{Pure position feedback}
As mentioned in the previous section, direct velocity measurements (from which the momentum can be computed) are typically not available in mechanical (robotic) control systems. However, the momentum vector $p_k$ is needed in Eq. \eqref{eq:3.112b} to determine the stage values $q_{k+\frac{1}{2}}$ and $p_{k+\frac{1}{2}}$ for the feedback law \eqref{eq:3.111}. To get along with only position measurements, we propose the reconstruction of $p_k$ from the measured position $q_k$ and its adjacent stage values $q_{k\pm\frac{1}{2}}$. The resulting discrete-time system has a beautiful interpretation in terms of one of the most prominent symplectic integration schemes.

\subsubsection{Potential energy shaping, constant mass matrix}
We start again with the case of only potential energy shaping, and assume this time first a constant mass matrix $M = const.$

\begin{thm}
	We consider the implicit control law \eqref{eq:3.090} with $q_{k+\frac{1}{2}}$ (and $p_{k+\frac{1}{2}}$) determined from the solution of the system of equations \eqref{eq:3.062}. In addition, we reconstruct the necessary momentum value in $t_k = kh$ via
	\begin{equation}
		\label{eq:3.170}
		p_{k} = M \frac{q_{k+\frac{1}{2}} - q_{k-\frac{1}{2}}}{h}.
	\end{equation}
	The value $q_{k}$ is available from measurement, and $q_{k-\frac{1}{2}}$ is a stored previous stage value. Then, (a) the resulting implicit feedback control \eqref{eq:3.090} depends only on position measurements and (b) the closed-loop system is the St{\"o}rmer-Verlet discretization of the continuous-time target system \eqref{eq:3.070} with $M = const.$
\end{thm}

\begin{pf}
	Statement (a) is evident from the fact that $p_k$ is replaced by an expression, which depends on $q_k$ and the stage values $q_{k-\frac{1}{2}}$, $q_{k+\frac{1}{2}}$ that are computed and stored in every time step. To prove statement (b), we consider the resulting momentum update equation \eqref{eq:3.060b} for $M = const.$ and $\nabla_q H_d = \nabla V_d$. We substitute $p_k$ and $p_{k+1}$ according to \eqref{eq:3.170} and obtain
	\begin{equation}
		M \frac{q_{k+\frac{3}{2}} - q_{k+\frac{1}{2}}}{h} - 
		M \frac{q_{k+\frac{1}{2}} - q_{k-\frac{1}{2}}}{h} +
		h \nabla V_d(q_{k+\frac{1}{2}}) = 0.
	\end{equation}
	Multiplying with $h \bar M$, this equation boils down to
	\begin{equation}	
		\label{eq:3.180}
		q_{k+\frac{3}{2}} - 2 q_{k+\frac{1}{2}} + q_{k-\frac{1}{2}} = - h^2 \bar M \nabla_q V_d (q_{k+\frac{1}{2}}),
	\end{equation}
	which is the two-step formulation \eqref{eq:20.30.70} of the St{\"or}mer-Verlet scheme on the staggered grid of position stage values.
	\hfill \quad \qed
\end{pf}

\begin{rem}
	Combining reconstruction of the momenta in the sampling instants and the discrete-time target dynamics, we obtain 
	\begin{subequations}
	\label{eq:3.190}
	\begin{align}
		\label{eq:3.190a}	
		q_{k+\frac{1}{2}} &= q_{k-\frac{1}{2}} + h \bar M p_k\\
		\label{eq:3.190b}
		p_{k+1} &= p_k - h \nabla V_d(q_{k+\frac{1}{2}}),
	\end{align}
	\end{subequations}
	which corresponds to the one-step formulation \eqref{eq:20.30.80}, shifted by $\frac{h}{2}$.
\end{rem}

\begin{rem}
	The following question can be posed. Does \eqref{eq:3.190a} contradict the (unchangeable by control) description of the coordinate update in the approximate sampling model \eqref{eq:3.030a}? The answer is no. While the latter describes the sampling process, the former states a relation between the \emph{stage values}, which are not measured, but which are computation quantities, and which are used to reconstruct the momenta in the sampling instants.
\end{rem}

\subsubsection{The general case}
We now apply the the reconstruction of momenta \eqref{eq:3.170} for state feedback to the general class of target systems 
\begin{equation}
	\label{eq:3.200}
	\dot q(t) = a(q(t), p(t)), \qquad 
	\dot p(t) = b_d(q(t), p(t))
\end{equation}
with the shortcut $a(q(t), p(t)) = \bar M(q(t)) p(t)$ and the subscript ``d'' for desired, indicating that only the momentum differential equation is modified by control.  In particular, we describe, for compactness of notation, the generalized Hamiltonian dynamics \eqref{eq:3.072} in this form.

\begin{thm}
	The control law \eqref{eq:3.111}, with $p_{k+\frac{1}{2}}$ and $q_{k+\frac{1}{2}}$ solutions of \eqref{eq:3.112a} and \eqref{eq:3.112b} and the reconstruction of the momenta in the sampling instants
		\begin{equation}
			\label{eq:3.205}
			p_{k} = M(q_k) \frac{q_{k+\frac{1}{2}} - q_{k-\frac{1}{2}}}{h}
		\end{equation}  
		leads to a closed-loop system, which is a second order approximation of the St{\"o}rmer-Verlet discretization of \eqref{eq:3.072}.
\end{thm}

\begin{pf}
	The St{\"o}rmer-Verlet discretization of \eqref{eq:3.200} on staggered grids has the form
	\begin{subequations}
	\label{eq:3.210}
		\begin{align}
			\label{eq:3.210a}
			q_{k+\frac{1}{2}} &= q_{k-\frac{1}{2}} + \frac{h}{2} \left ( a(q_{k-\frac{1}{2}}, p_k) + a(q_{k+\frac{1}{2}}, p_k) \right)\\
			\label{eq:3.210b}
			p_{k+1} &= p_{k} + \frac{h}{2} \left ( b_d(q_{k+\frac{1}{2}}, p_k) + b_d(q_{k+\frac{1}{2}}, p_{k+1}) \right),
		\end{align}
	\end{subequations}
	see \cite{hairer2003geometric}, Section 1.8, which refers to the original unpublished work \cite{devogelaere1956methods}.
	In contrast to that, the target momentum equation \eqref{eq:3.110b} and Eq. \eqref{eq:3.205} to reconstruct the momenta can be rearranged in the form
	\begin{subequations}
	\label{eq:3.220}
		\begin{align}
			\label{eq:3.220a}
			q_{k+\frac{1}{2}} &= q_{k-\frac{1}{2}} + h a(q_{k}, p_k)\\
			\label{eq:3.220b}
			p_{k+1} &= p_{k} + h b_d(q_{k+\frac{1}{2}}, p_{k+\frac{1}{2}}).
		\end{align}
	\end{subequations}	
	With 
	\begin{align}
		\label{eq:3.222}
		&a(q_{k-\frac{1}{2}}, p_k) + a(q_{k+\frac{1}{2}}, p_k) =\\ \nonumber
		&2 a(q_k, p_k) + \left. \frac{\partial a}{\partial q} \right|_{(q_k, p_k)} \left( (- \frac{h}{2} \dot q_k + o(h)) + (\frac{h}{2} \dot q_k + o(h) )  \right)
	\end{align}
	we verify that \eqref{eq:3.210a} and \eqref{eq:3.220a} differ by an expression of polynomial degree greater than 2 in $h$. The same holds for \eqref{eq:3.210b} and \eqref{eq:3.220b}.
	\hfill \quad \qed
\end{pf}

\begin{rem}
	It is possible to use the momentum equation \eqref{eq:3.210b} for the target system instead of \eqref{eq:3.110b} and to modify the reconstruction equation for the momenta as follows,
	\begin{equation}
		\frac{1}{2} \left( \bar M(q_{k-\frac{1}{2}}) + \bar{M}(q_{k+\frac{1}{2}}) \right) p_k = \frac{q_{k+\frac{1}{2}} - q_{k-\frac{1}{2}}}{h},
	\end{equation}		
	in order to have \eqref{eq:3.210a} for the update of the coordinates. For the following practical reasons, we consider it as legitimate to work with the approximation of second order: (i) The future value $p_{k+1}$ appears in \eqref{eq:3.210b} as the argument of a nonlinear function and an additional unknown which has to be extrapolated from given data, e.g. $p_{k+\frac{1}{2}}$. (ii) The quality of the results presented in the next section confirms the validity of this approximation, and suggests that the increased complexity and computational effort to achieve \eqref{eq:3.210} are not justified.
\end{rem}

\subsection{Lagrangian perspective}
We now take the Lagrangian perspective, which can be more convenient for implementation, as the velocities are directly measured or reconstructed from the positions without involvement of the mass matrix. We depart from the implicit midpoint rule discretization of \eqref{eq:20.30.41}.

\begin{cor}
	\label{cor:lagrangian}
	The discrete-time model
	\begin{subequations}
		\label{eq:4.160}
		\begin{align}
		\label{eq:4.160a}
		q_{k+1} &= q_{k} + h v_{k+\frac{1}{2}}\\
		\label{eq:4.160b}
		v_{k+1} &= v_k + hf(q_{k+\frac{1}{2}},v_{k+\frac{1}{2}}) + h\bar M(q_{k+\frac{1}{2}})u_k
		\end{align}
	\end{subequations}
	is a second order approximation of the sampled mechanical system.	
\end{cor}

Under the assumption of piecewise constant inputs, the proof is similar to the one of Theorem \ref{thm:01}.

\begin{dfn}
	For the -- possibly time-varying -- continuous-time target system \eqref{eq:20.30.43}
	the discrete-time approximation with the implicit midpoint rule is
	\begin{subequations}
		\label{eq:4.180}
		\begin{align}
		\label{eq:4.180a}
		q_{k+1} &= q_k + h v_{k+\frac{1}{2}}\\
		\label{eq:4.180b}
		v_{k+1} &= v_k + hf_d(q_{k+\frac{1}{2}},v_{k+\frac{1}{2}},t_{k+\frac{1}{2}})
		\end{align}
	\end{subequations}	 
	with $t_{k+\frac{1}{2}} = t_k + \frac{h}{2}$ and the stage values
	\begin{equation}
		q_{k+\frac{1}{2}} = \frac{q_k+ q_{k+1}}{2}, \qquad
		v_{k+\frac{1}{2}} = \frac{v_k+ v_{k+1}}{2}.
	\end{equation}
\end{dfn}

With \eqref{eq:4.180a} and the definition of the stage value $q_{k+\frac{1}{2}}$, we can express $v_{k+\frac{1}{2}}$ as follows:
\begin{equation}
	v_{k+\frac{1}{2}} = \frac{2}{h} (q_{k+\frac{1}{2}} - q_k),
\end{equation}
which allows us to state the following corollary.

\begin{cor}
	\label{thm:lagrangian}
	The state feedback
	\begin{equation}
		\label{eq:4.200}
		u_k = M(q_{k+\frac{1}{2}})\left(f_d(q_{k+\frac{1}{2}},v_{k+\frac{1}{2}}, t_{k+\frac{1}{2}}) - f(q_{k+\frac{1}{2}},v_{k+\frac{1}{2}})\right)
	\end{equation}	
	transforms \eqref{eq:4.160} into the target system \eqref{eq:4.180}. The stage values $q_{k+\frac{1}{2}}$ and $v_{k+\frac{1}{2}}$ are solutions of the system of equations
	\begin{subequations}
		\label{eq:4.202}
	\begin{align}
		q_{k+\frac{1}{2}}  &= q_k + \frac{h}{2} v_{k+\frac{1}{2}}\\
		v_{k+\frac{1}{2}}  &= v_k + \frac{h}{2} f_d(q_{k+\frac{1}{2}}, v_{k+\frac{1}{2}}, t_{k+\frac{1}{2}}).
	\end{align}
	\end{subequations}
	For a pure position feedback, the velocities are approximated by
	\begin{equation}
		\label{eq:4.204}	
		v_k = \frac{q_{k+\frac{1}{2}} - q_{k-\frac{1}{2}}}{h},
	\end{equation}
	which again yields an interpretation of the resulting target system in terms of the St{\"o}rmer-Verlet integration scheme on staggered grids.
\end{cor}

\begin{pf}
	Inserting \eqref{eq:4.204} and the corresponding expression for $v_{k+1}$ in \eqref{eq:4.180b} yields a second order centered finite difference approximation of $\ddot q(t) = f_d(q(t), \dot q(t), t)$ in $t_{k+\frac{1}{2}}$. If $f_d(q(t), \dot q(t), t) = f_d(q(t))$, this corresponds exactly to the St{\"o}rmer-Verlet scheme for a separable Hamiltonian system on the staggered grid.
	\hfill \quad \qed	
\end{pf}

For the implementation of the described implicit discrete-time control scheme, note that an initial value $v_0$ (typically zero) is necessary in the first step. In all other sampling intervals, the stored value $q_{k-\frac{1}{2}}$ is used for the solution of the system of equations \eqref{eq:4.202}, \eqref{eq:4.204}.

\section{Numerical experiments}
\label{sec:4}
The effectiveness of our approach is now illustrated with three benchmark examples for fully actuated mechanical systems. The first two examples are presented in the (port-)Hamiltonian frame, while the Lagrangian perspective is assumed in the third example.

\subsection{Simulation settings}	
For all simulation experiments we used \emph{Matlab}/\emph{Simulink} and the following settings. The continuous-time open-loop system is integrated with the solver \emph{ode23} and a relative tolerance of $10^{-6}$. Sample and zero-order hold according to Fig. \ref{fig:sampled-control-loop} are emulated with the corresponding Simulink blocks. Nonlinear systems of equations to determine the stage values of coordinates and momenta/velocities are solved numerically with \emph{fsolve}. A rough estimation of the computation time (using the commands \emph{tic} and \emph{toc}) shows that the equations are solved in around $5\:\text{ms}$ on a sub-standard laptop PC.

\subsection{Mass-spring system}
High stiffness(es) of the virtual spring(s) in impedance control lead to oscillations and loss of stability at high sampling times. On the basic example of a mass-spring system, we show that the maximum assignable stiffness can be drastically increased when applying our symplectic discrete-time control design.

The considered system with $n=1$ degree of freedom can be written in the form \eqref{eq:20.30.10} with Hamiltonian
$H(q,p) = \frac{1}{2}kq^2 + \frac{1}{2}\frac{p^2}{m}$
and the constant parameters $m=1\:\text{kg}$, $k=0.5\:\text{N/m}$. The target system
	\begin{equation}
		\label{eq:5.030}
		\bmat{\dot q(t)\\ \dot p(t)}
		=
		\bmat{0 & 1\\ - 1 & -d }
		\bmat{ c q(t) \\ \frac{p(t)}{m} }
	\end{equation}
has the desired Hamiltonian (only potential energy shaping)
$H_d(q,p) = \frac{1}{2}cq^2 + \frac{1}{2}\frac{p^2}{m}$,
with a new virtual stiffness $c>k$ and damping coefficient $d=0.1\:\text{Ns/m}$. By linearity of the target system, the equations \eqref{eq:3.062} can be solved analytically to determine the \emph{symplectic} control law
	\begin{equation}
	\label{eq:5.045}
		u_{k} = (k-c)q_{k+\frac{1}{2}}-\frac{d}{m}p_{k+\frac{1}{2}}.
	\end{equation}
For comparison, we also consider the \emph{quasi-continuous} control law
	\begin{equation}
	\label{eq:5.050}
		u_{k} = (k-c)q_k-\frac{d}{m}p_k.
	\end{equation}
	
\begin{figure}[htbp]
	\centering
	\includegraphics{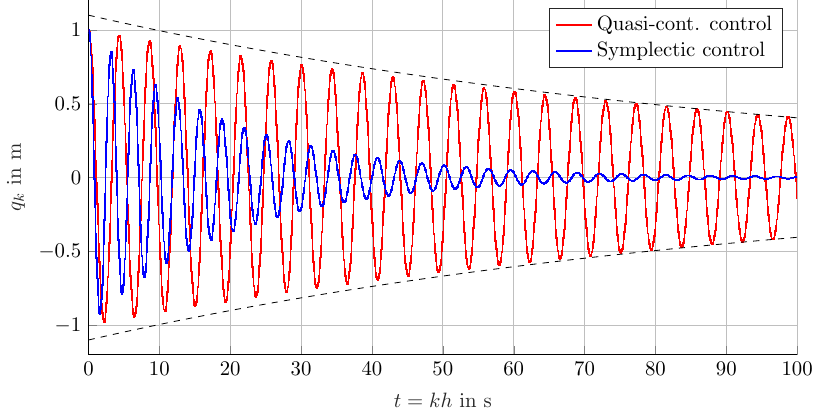}
	\caption{Initial value responses $q_k$ with maximum achievable stiffness $c_{\mathrm{max}}$ for  $h=0.1\:\text{s}$ (see Fig. \ref{fig:5.011}). Not only is $c_{\mathrm{max}}$ higher under symplectic control, but also the $L_2$ norm $\| q_k \|_h$ is smaller for input signals with identical energy (see Fig. \ref{fig:5.012}).}
	\label{fig:5.02}
\end{figure}

\begin{figure}[htbp]
	\centering
	\includegraphics{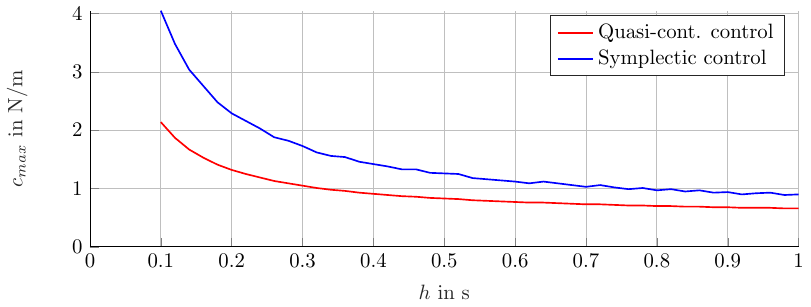}
	\caption{Maximum achievable stiffnesses $c_{\mathrm{max}}$ with quasi-continuous and symplectic control and identical input energy \eqref{eq:5.054}.}
	\label{fig:5.011}
\end{figure}

\begin{figure}[htbp]
	\centering
	\includegraphics{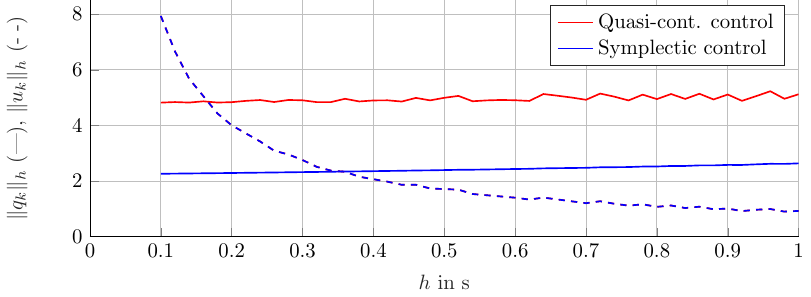}
	\caption{With the same input energy $\| u_k \|_h$, the norm of the coordinate (error) $\| q_k \|_h$ is half as big under symplectic control. In either of the cases, the corresponding maximum stiffness $c_{max}$ is assigned.}
	\label{fig:5.012}
\end{figure}

The maximum assignable stiffness $c_{\mathrm{max}}$ under both controllers is determined as follows. 
For every sampling time $h$, the limit $c_{\mathrm{max}}$ under the quasi-continuous controller is determined based on the condition\footnote{$\|x_k\|_{\infty}$ is the maximum norm of a discrete-time signal $x_k$.}
\begin{equation}
	\label{eq:5.052}
	\| e^{\alpha k h} q_k \|_{\infty}  < 1.1 q_0, \qquad \alpha = 0.1\frac{d}{m},
\end{equation}
which means that the coordinate remains confined to the dashed tube in Fig. \ref{fig:5.02}. The initial conditions are  $q(0)=1\:\text{m}$ and $\dot{q}(0)=0\:\text{m/s}$. The red curve in Fig. \ref{fig:5.011} shows the achievable stiffnesses. At the same time the \emph{discrete-time} $L_2$ norm\footnote{See \cite{iserles2009first}, Section 16.2. The norm tends to the continuous $L_2$ norm for $h\rightarrow 0$.} of the input sequence $u_k$ on the time interval $[0, Nh]$
	\begin{equation}
		\label{eq:5.054}
		\| u_k \|_h = \left( h \sum_{k=0}^N \|u_k\|^2 \right)^{\frac{1}{2}}
	\end{equation}
is stored. The maximum stiffness under the symplectic controller is now determined also based on the criterion \eqref{eq:5.052}, but under the additional constraint that the $L_2$ norm of the input does not exceed the maximum value for the quasi-continuous case. Figure \ref{fig:5.012} shows the norms of the control signal and the position for the maximum assigned stiffness in either of the cases.

\subsection{Two link robot arm, PD controller}
\begin{figure}[htbp]
	\centering
	\includegraphics{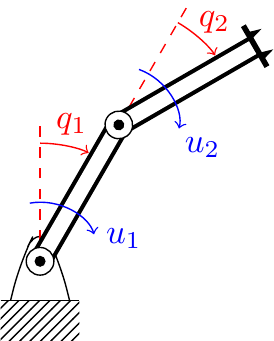}
	\caption{Two link robot arm}
	\label{fig:5.03}
\end{figure}

In this and the following subsection, we consider the two link robot arm depicted in Fig. \ref{fig:5.03}. The \mbox{(port-)}Hamiltonian model has the form \eqref{eq:20.30.10}, \eqref{eq:20.30.20} with mass matrix
	\begin{equation}
		\label{eq:5.060}
		M(q_2) = \bmat{c_1 + c_2 + 2c_3\cos(q_2) & c_2 + c_3\cos(q_2) \\ c_2 + c_3\cos(q_2) & c_2 }
	\end{equation}
and potential energy
	\begin{equation}
		\label{eq:5.070}
		V(q_1,q_2) = c_4g\cos(q_1) +  c_5g\cos(q_1+q_2).
	\end{equation}
The constants $c_1,\dots,c_5$ include the arm masses $m_i$, the arm inertias $J_i$ around the centers of gravity, the total arm lengths $L_i$, the distances $l_i$ from the joints to the centers of gravity and a motor mass $m_M$ placed at the second joint:
	\begin{equation}
		\label{eq:5.080}
		\begin{alignedat}{3}
		c_1 &= J_1 + m_1l_{1}^{2} + (m_M+m_2)L_{1}^{2}, & &\\ 
		c_2 &= J_2 + m_2l_{2}^{2}, & c_3 &= m_2L_1l_2, \\ 
		c_4 &= m_1l_1 + (m_M+m_2)L_1, & c_5 &= m_2l_2.
		\end{alignedat}
	\end{equation}
The parameters for the two links are $m_{1}=m_{2}=0.885\: \text{kg}$, $J_{1}=J_{2}=3.27\cdot10^{-3}\: \text{kgm}^{2}$, $L_{1}=L_{2}=0.2\: \text{m}$, $l_{1}=l_{2}=0.1\: \text{m}$, the motor (stator) mass is $m_M = 1.0\: \text{kg}$, and $g=9.81\: \text{m/s}^{2}$.

In this simulation study, we consider a PD controller with gravity compensation according to Subsection \ref{subsec:PD}. to stabilize the upright position $q_d = [0\;\;0]^T$ from the initial configuration $q(0) = [\pi \;\; 0]^T$ at rest. The assigned stiffness and damping matrices are $K = \mathrm{diag}\{0.1, 0.1\}\:\text{N/m}$ and $D = \mathrm{diag}\{0.1, 0.1\}\:\text{Ns/m}$. We compare the closed-loop dynamics under the symplectic implicit control law
	\begin{equation}
		\label{eq:5.110}
		u_k = \nabla V(q_{k+\frac{1}{2}}) - D\dot{q}_{k+\frac{1}{2}} - Kq_{k+\frac{1}{2}}
	\end{equation}	
with the quasi-continuous implementation
	\begin{equation}
		\label{eq:5.115}
		u_k = \nabla V(q_k) - D\dot{q}_k - Kq_k.
\end{equation}
The implicit control law \eqref{eq:5.110} becomes a pure position feedback by the reconstruction of the velocity in the sampling instants according to \eqref{eq:4.204}.

\begin{figure}[htbp]
	\centering
	\begin{subfigure}{\linewidth}
		\includegraphics{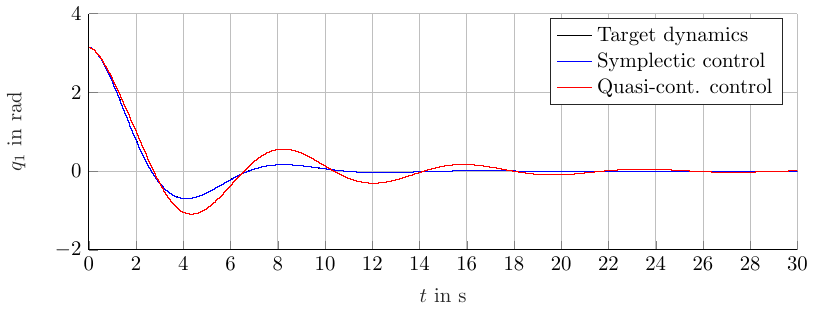}
		\includegraphics{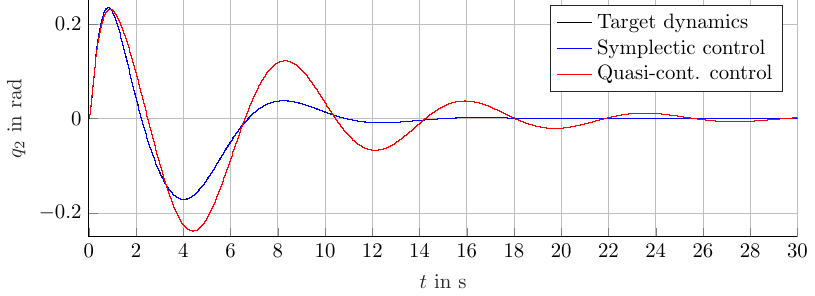}
		\caption{Joint angles of the target dynamics (black), under the symplectic controller \eqref{eq:5.110} (blue) and with the quasi-continuous implementation of the control law \eqref{eq:5.115} (red).}
		\label{fig:5.041}
	\end{subfigure}
	\\
	\vspace*{10 pt}
	\begin{subfigure}{\linewidth}
		\includegraphics{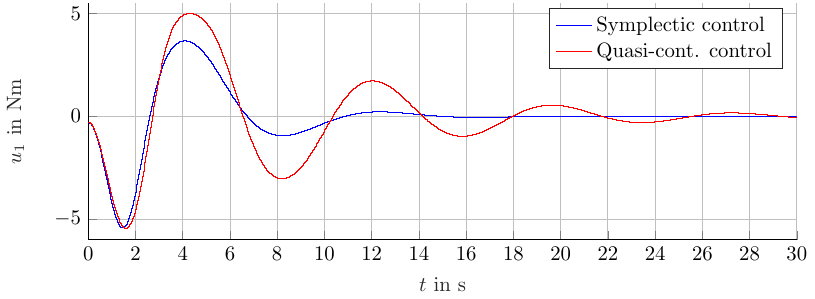}
		\includegraphics{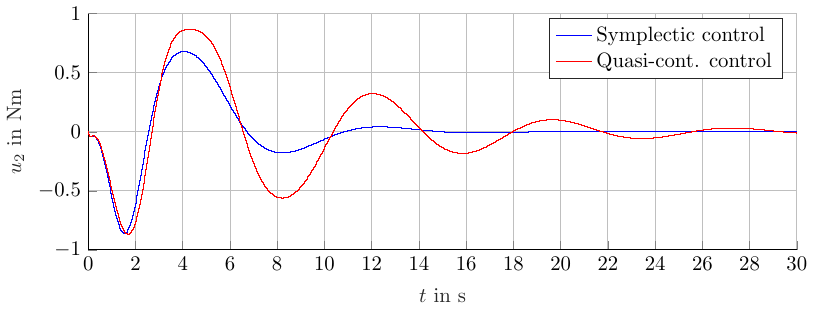}
		\caption{Joint torques for symplectic (blue) and quasi-continuous (red) control.}
		\label{fig:5.042}
	\end{subfigure}
	\caption{PD robot control with $h=0.02\:\text{s}$}
	\label{fig:5.04}
\end{figure}

\begin{figure}[htbp]
	\centering
	\begin{subfigure}{\linewidth}
		\includegraphics{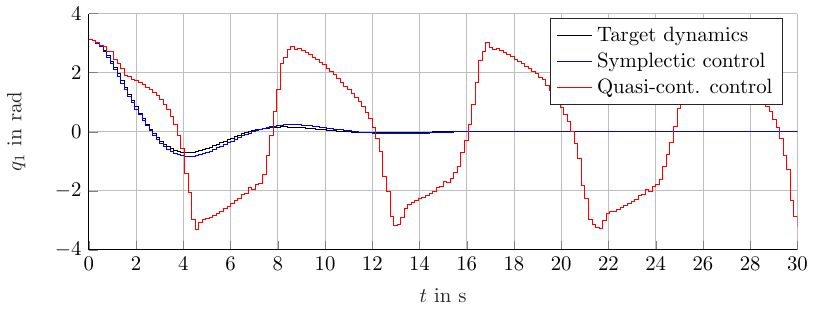}
		\includegraphics{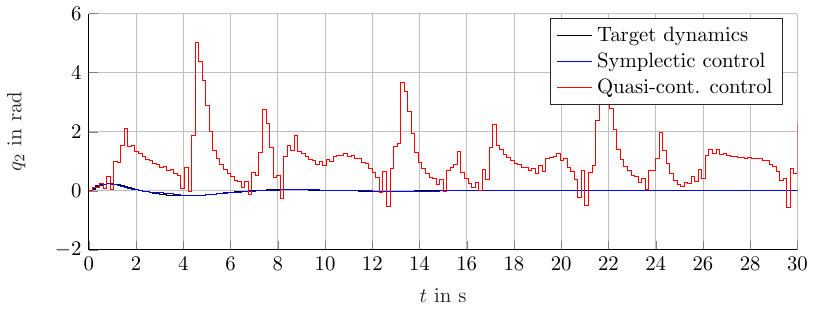}
		\caption{Joint angles.}
		\label{fig:5.061}
	\end{subfigure}
	\\
	\vspace*{10 pt}
	\begin{subfigure}{\linewidth}
		\includegraphics{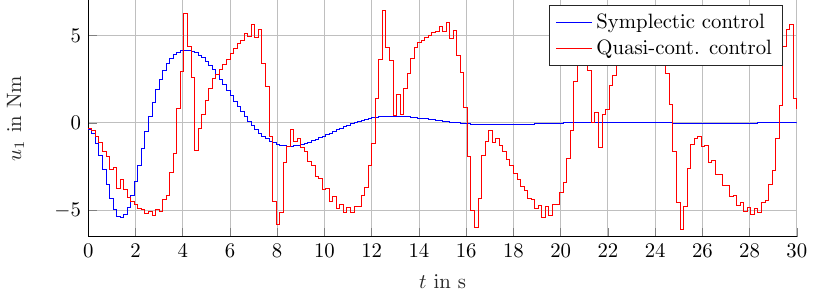}
		\includegraphics{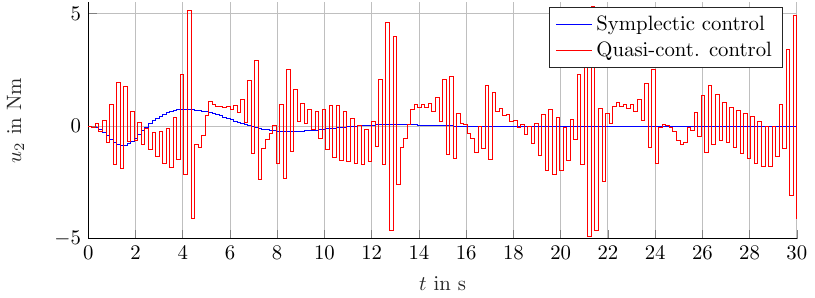}		
		\caption{Joint torques.}
		\label{fig:5.062}
	\end{subfigure}
	\caption{PD robot control with $h=0.15\:\text{s}$}
	\label{fig:5.06}
\end{figure}

The results of this simulation study are presented in Fig. \ref{fig:5.04} for $h= 0.02\: \text{s}$ and in Fig. \ref{fig:5.06} for $h = 0.15\: \text{s}$. At the short sampling time, the symplectic controller perfectly matches the prescribed behavior of the target system (blue and black curve in Fig. \ref{fig:5.041}) while the quasi-continuous implementation already shows a remarkable deviation (red curve) from the desired dynamics, see also Fig. \ref{fig:5.042} for the input signals. Even more impressive is the situation for $h=0.15\: \text{s}$, where the quasi-continuous control completely fails, while the symplectic discrete-time controller produces only a slight deviation from the target dynamics.

\subsection{Two link robot arm, computed torque}
Finally, we place a (virtual) massless pencil at the tool center point (TCP), whose Cartesian coordinates are
	\begin{equation}
		\xi = \bmat{x_{TCP} \\ y_{TCP}} = \bmat{L_1\sin(q_1) + L_2\sin(q_1+q_2) \\ L_1\cos(q_1) + L_2\cos(q_1+q_2)}.
	\end{equation}
The robot is supposed to draw a circle in the Cartesian task space, according to the reference trajectory\footnote{This would be a path following problem. In the given context, we simply consider the trajectory tracking problem, without using dynamic feedforward control.}
	\begin{equation}
		\xi_d(t)=\bmat{L_1 + \frac{L_2}{2}\cos(\Omega t) \\ L_1 + \frac{L_2}{2}\sin(\Omega t)}
	\end{equation}
with $\Omega = 0.1\:\text{rad/s}$. Using the inverse kinematics, the desired trajectory in the joint space $q_d(t)$ and the velocities $\dot q_d(t)$ can be computed, which allows to express a \emph{computed torque} control law according to Section \ref{subsec:computed-torque}. With the constant diagonal design matrices $M_d = \mathrm{diag}\{0.1, 0.013\}\:\text{kgm}^2$, $K = \mathrm{diag}\{0.3, 0.03\}\:\text{N/rad}$ and $D = \mathrm{diag} \{ 0.3, 0.03 \} \:\text{Ns/rad}$, we obtain the quasi-continuous control law	 
	\begin{equation}
		\label{eq:5.140}
		\begin{split}
			u_k &= C(q_k,\dot{q}_k)\dot{q}_k + \nabla V(q_k) + M(q_k)\ddot{q}_d(t_k) \\
			&  + M(q_k)M_d^{-1}(-K(q_k-q_d(t_k))-D(\dot{q}_k-\dot{q}_d(t_k))).
		\end{split}
	\end{equation}
The symplectic discrete-time control law according to \eqref{eq:4.200} follows from replacing the index $k$ with $k+\frac{1}{2}$ in the terms on the right hand side. As in the previous example, no velocity measurement in the sampling instants is required, as the system of equations \eqref{eq:4.202}, with reconstruction of the velocity \eqref{eq:4.204} using the position stage values, is solved numerically in every sampling interval.
	
\begin{figure}[htbp]
	\centering
	\includegraphics{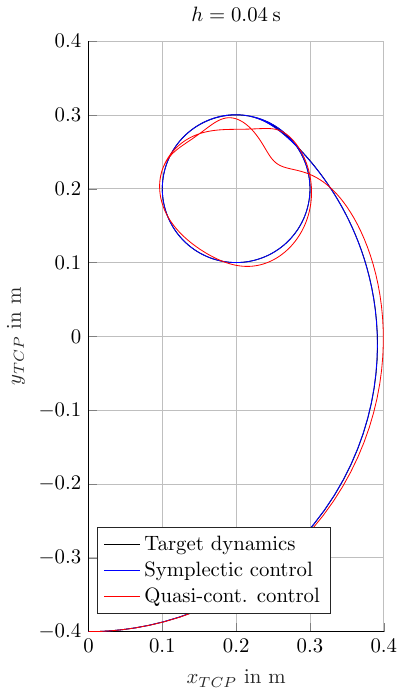}
	\includegraphics{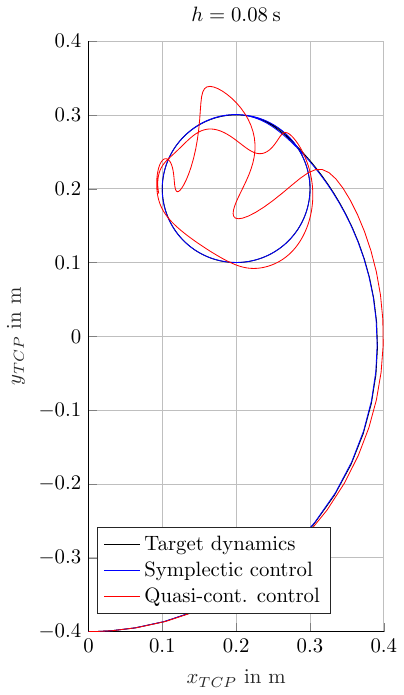}
	\caption{For $h=0.04\:\text{s}$ (left) the TCP almost perfectly follows the desired circular trajectory (black: trajectory of the target system) under the symplectic control law (blue), while the performance of the quasi-continuous implementation (red) is not satisfactory. A comparable situation with a useless red trajectory is shown for  $h=0.08\:\text{s}$ on the right. Even for $h= 0.15\:\text{s}$ (not depicted) the deviations of the symplectic control from the reference are relatively small, while the behavior of the quasi-continuous control is further deteriorated.}
	\label{fig:5.05}
\end{figure}

Exemplary simulation results are depicted in Fig. \ref{fig:5.05}, which shows phase diagrams of the TCP for different sampling times.

\section{Conclusions}
\label{sec:5}
We presented a novel, systematic approach for discrete-time control design of fully actuated mechanical systems. Our approach is motivated by energy shaping as an intermediate step (before damping injection/asymptotic stabilization), where desired, conservative dynamical behavior is imposed to the control system. Symplectic integration with its conservation properties, in particular the modified Hamiltonian, is a conceptually straighforward vehicle to transfer energy shaping arguments to discrete time.

Departing from a second order accurate description of the sampling process by the implicit midpoint rule, we proposed implicit state feedback control laws for different control tasks by defining the corresponding target systems via the implicit midpoint rule as well. As a particularly beautiful feature of the design approach, we get rid of velocity measurements by a simple (trapezoidal rule) reconstruction of the velocities/momenta, and obtain a closed-loop discrete-time system, which has a clear interpretation in terms of the famous St{\"o}rmer-Verlet integration scheme and its variation for general partitioned systems.

Simulations of three benchmark examples give clear evidence of the utility of the design method, which removes deteriorating effects of sampling on stability and performance, even at very low sampling rates.

We are currently working on the application of our approach to serial elastic manipulators, i.e., the underactuated case, including the experimental validation with flexible light weight robots. The presented approach is -- by the initial use of the implicit midpoint rule -- not restricted to mechanical systems. It can be also applied to further, different classes of nonlinear control problems. We are also interested in exploiting different types of sampling mechanisms and higher order approximation schemes.

\begin{ack}
	The work was supported by Deutsche Forschungsgemeinschaft (project number 317092854) and Agence Nationale de la Recherche (ID ANR-16-CE92-0028), project INFIDHEM. 
\end{ack}

\bibliographystyle{plain}
\bibliography{2020_koty_thom_automatica_v1}

\end{document}